\documentclass[12pt]{article}
\begin{document}
\begin{center}
Space-Time Symmetries in Noncommutative Gauge Theory:\\
A Hamiltonian Analysis
\vskip 1cm
Subir Ghosh{\footnote{e-mail:sghosh@isical.ac.in}}\\
\vskip 1cm
Physics and Applied Mathematics Unit,\\
Indian Statistical Institute,\\
203 B. T. Road, Kolkata 700108, \\
India.
\end{center}
\vskip 3cm
{\bf Abstract:}\\
We study space-time symmetries in Non-Commutative (NC) gauge
theory in the (constrained) Hamiltonian framework. The specific example of NC
$CP(1)$ model, posited in \cite{sg}, has been considered. Subtle
features of   Lorentz invariance violation in  NC field theory
were pointed out in \cite{har}. Out of the two - Observer and 
Particle - distinct types of Lorentz transformations, symmetry
under the former, (due to the translation invariance), is
reflected in the conservation of energy and momentum in  NC
theory. The constant tensor $\theta_{\mu\nu}$ (the
noncommutativity parameter) destroys invariance under the latter.

In this paper we have constructed
 the Hamiltonian and momentum operators which are the
 generators of time and space translations respectively. This is related to the Observer Lorentz invariance. We have also shown that the Schwinger condition and subsequently 
 the Poincare algebra is not obeyed and that one can not derive a Lorentz covariant dynamical
 field equation. These features signal a loss of the Particle Lorentz symmetry.
The basic observations in the present work will be relevant in the Hamiltonian
study of a generic noncommutative field theory.

\newpage

\begin{center}
{\bf{Introduction}}
\end{center}

In recent years, Non-Commutative (NC) field theories have  become the focus of intense
research activity after its connection to low energy string
physics was elucidated by Seiberg and Witten \cite{sw,rev}.
Specifically, the open string boundaries, attached to D-branes
\cite{pol}, in the presence of a two-form background field, turn
into NC spacetime coordinates \cite{sw}. (This phenomenon has been
recovered from various computational schemes \cite{others}.) The
noncommutativity induces an NC D-brane world volume and hence
field theories on the brane become NC field theories.

NC   field theories have revealed unexpected textures in the
conventional field theory framework - UV-IR mixing  \cite{min1},
soliton solutions in higher dimensional scalar theories
\cite{min2}, dipole-like elementary charged excitations
\cite{jab}, etc. are some of them. The inherent non-locality, (or
equivalently the introduction of a length scale by
$\theta^{\rho\sigma}$ - the noncommutativity parameter), of the NC
field theory is manifested through these peculiar properties that
are absent in the corresponding ordinary spacetime theories.

Solitons in NC $CP(1)$ model have been found
\cite{nccp}, very much in analogy to their counterpart in ordinary
spacetime.  In this framework, NC theories are treated as
systems of operator  valued fields and one works  directly  with
operators on the quantum phase space, characterized by the
noncommutativity condition
\begin{equation}
[x^{\rho},x^{\sigma}]_{*}=i\theta^{\rho\sigma}.
\label{nc}
\end{equation}
With only spatial noncommutativity on the NC plane, the above
simplifies to a Heisenberg algebra,
$[x^1,x^2]_{*}=x^{1}*x^{2}-x^{2}*x^{1}=i\theta
^{12}=i\epsilon^{12}\theta =i\theta $ which in the complex
coordinates reduces to the creation annihilation operator algebra
for the simple Harmonic Oscillator. Thus to a function in the NC
spacetime, through Weyl transform, one associates an operator
acting on the Hilbert space, in a basis of a simple Harmonic
Oscillator eigenstates. Explicit details of the computations in
this particular model are found in \cite{nccp}.

In the present work, we will concentrate on a specific NC gauge
theory, that was recently proposed \cite{sg} as  an alternative
formulation of the NC extension of the $CP(1)$ model. Due to the
presence of the $U(1)$ invariance, (induced by the $CP$
variables), the Seiberg-Witten map \cite{sw} plays a pivotal role
in our scheme. It is used to convert the NC to a theory expressed
in terms of ordinary fields, with noncommutative effects appearing
as $\theta$-dependent interaction terms. We found in \cite{sg} that our model allows solutions obeying  a
(Bogolmolny)  lower bound in energy, (protected by topological
considerations), which are  the solitons
of the NC $CP(1)$ model. We also noted in  \cite{sg} that (unlike the commutative case) additional restrictions on the $CP(1)$ variables appear when the BPS equations are considered as a subset of the variational equation of motion. The reason might be the perturbative (in $\theta $) nature of the formalism. In fact it is well known that there are complications in the
definition of the Energy-Momentum  (EM) tensor in NC field theory
\cite{grim}.

The EM tensor in a generic NC field theory has been discussed in 
 \cite{grim,also} in the Lagrangian framework. Hamiltonian analysis of NC theories have been performed in \cite{ham}. The novel feature of the present work is the study of the Poincare algebra, leading to the Lorentz invariance violation.

Of special interest will be an explicit demonstration of the
validity of the ideas introduced in \cite{har}, in the context of
Lorentz symmetry violation in NC field theories. The issue is
subtle since there exists \cite{coll} {\it{two}} distinct types of
Lorentz transformations: Observer and Particle Lorentz
transformations. The NC action, (as well as the fields and the
constant tensor $\theta_{\mu\nu}$ comprising it), transform
covariantly under the former, thereby yielding conservation of
energy and momentum, at least when only spatial noncommutativity
is present. This is expected due to the translation invariance of
the theory. On the other hand, the essential ingredients of a
relativistic theory - the Schwinger condition \cite{schw} and subsequent
Poincare algebra - are not respected. This indicates a loss of the
Particle Lorentz symmetry. These generic features will emerge
naturally in our Hamiltonian formulation. We have restricted the discussion to spatial noncommutativity only so that a conventional Hamiltonian analysis can be carried through.

Let us put our work in its proper perspective. This paper is a
sequel to \cite{sg} where we provide a field theoretic analysis of
the $CP(1)$ model in a Hamiltonian framework, keeping in mind the
future possibility of quantization of the model. In particular, we
study in detail the nature (of the violation) of the Poincare
algebra. This is probably the first example of an in depth study
of a specific NC field theory model in the Hamiltonian scheme,
enunciated by Dirac \cite{di}. Our analysis reveals both expected
and unexpected features of the model. Some of the results are
generic to any NC field theory and some are specific to the
(spacetime) dimensionality of the problem. With its non-trivial
but simpler structure, 2+1-dimensional NC field theories can
become successful laboratories for higher dimensional studies, as
the present work indicates.

Furthermore, our analysis goes on to show that the alternative
definition of the NC $CP(1)$ model that we have posited in
\cite{sg}, leads to a consistent and well defined NC gauge theory,
which conforms to the expected features of such a system.

The paper is organized as follows: Section II introduces the
noncommutative spacetime and provides a brief digression of the NC
$CP(1)$ model \cite{sg}. The canonical EM tensor is studied in
Section III. Section IV discusses the Hamiltonian formulation of
the model with the associated constraint analysis. It also
exhibits the transformation properties of the fields under gauge
and spacetime transformations and the Hamiltonian equations of
motion. Section V is devoted to the study of the Schwinger
condition and Poincare algebra. The major contributions of the
present work are in Sections IV and V. Section VI provides a
summary and conclusions.

\vskip .5cm
\begin{center}
{\bf{Section II: Noncommutative $CP(1)$ model - a brief digression}}
\end{center}
The $CP(1)$ model in ordinary spacetime is described by the  gauge
invariant action {\footnote {Since the scenario is classical,
hermitian conjugate operator $\phi^{\dag}$ is replaced by complex
conjugate $\phi^{*}$ and operator ordering ambiguities are not
taken in to account anywhere. Adjacent $\phi$-terms without any
Roman index are assumed to be summed.}},
\begin{equation}
S=\int d^{3}x~[(D^{\mu}\phi )^{*}D_{\mu}\phi +\Lambda (\phi^{*}\phi -1)],
\label{ac}
\end{equation}
where $D_{\mu}\phi_{a} =(\partial_{\mu}-iA_{\mu})\phi_{a}$ defines the
covariant derivative  and the multiplier $\Lambda$ enforces the $CP(1)$
constraint. The equation of motion for $A_{\mu}$ leads to the
identification,
\begin{equation}
A_{\mu}=-i\phi^{*}\partial_{\mu}\phi .
\label{nca1}
\end{equation}
The "gauge field" $A_{\mu}$ - being a dependent variable - can be removed from the action classically using (\ref{nca1}). The  infinitesimal gauge transformation of
the variables are,
\begin{equation}
\delta \phi_{a}^{*}=-i\lambda \phi_{a}^{*} ~;~~\delta \phi_{a}=i\lambda
\phi_{a}~; ~~\delta A_{\mu}=\partial_{\mu}\lambda . \label{gt}
\end{equation}

Let us now enter the noncommutative spacetime. In constructing the NC $CP(1)$ (or any generic)
model   the following steps are taken \cite{sg}:\\
(i) The appropriate NC field theory is constructed in terms of NC
analogue fields $(\hat \psi )$ of the fields $(\psi)$ with the
replacement of ordinary products of fields $(\psi \varphi )$, by
the Moyal-Weyl $*$-product $(\hat \psi *\hat \varphi )$,
\begin{equation}
\hat \psi(x)*\hat
\varphi(x)=e^{\frac{i}{2}\theta_{\mu\nu}\partial_{\sigma_{\mu}}\partial_{\xi_{\nu}}}\hat
\psi(x+\sigma )\hat \varphi(x+\xi )\mid_{\sigma =\xi =0} =\hat
\psi(x)\hat \varphi(x)+
\frac{i}{2}\theta^{\rho\sigma}\partial_{\rho}\hat
\psi(x)\partial_{\sigma}\hat \varphi(x)+~O(\theta^{2}). \label{mw}
\end{equation}
The {\it{ hatted}} variables are NC degrees of freedom. We take
$\theta^{\rho\sigma}$   to be a real constant antisymmetric
tensor, as is customary \cite{sw}, (but this need not always be
the case \cite{sny}). The NC spacetime (\ref{nc}) follows from the above
definition.

Note that the effects of spacetime noncommutativity has been
accounted for by the  introduction of the $*$-product.
For gauge theories the Seiberg-Witten Map \cite{sw} plays a crucial role in
connecting $\hat \phi (x) $ to $\phi (x)$. This formalism allows us to study the effects
of noncommutativity as $\theta^{\rho\sigma}$ dependent interaction terms in an ordinary
spacetime field theory format. This is the prescription we will follow.\\

The first task is
to generalize the scalar  gauge theory (\ref{ac}) to its NC
version, keeping in mind that the latter must be $*$-gauge
invariant. The NC action (without the $CP(1)$ constraint) is,
\begin{equation}
\hat S=\int d^{3}x~(\hat D^{\mu}\hat\phi )^{*}*\hat
D_{\mu}\hat\phi =\int d^{3}x~ (\hat D^{\mu}\hat\phi )^{*}\hat
D_{\mu}\hat\phi , \label{ncac}
\end{equation}
where the NC covariant derivative is defined as
$$\hat D_{\mu}\hat\phi_{a} =\partial_{\mu}\hat\phi_{a}-i\hat A_\mu*\hat\phi_{a} .$$
 The NC
action (\ref{ncac}) is invariant under the $*$-gauge
transformations,
\begin{equation}
\hat \delta \hat\phi_{a}^{*}=-i\hat\lambda *\hat\phi_{a}^{*}
~;~~\hat\delta  \hat\phi_{a}=i\hat\lambda *\hat\phi_{a}~;~~\hat\delta \hat
A_{\mu}=\partial_{\mu}\hat\lambda +i[\hat \lambda ,\hat
A_{\mu}]_{*}. \label{ncgt}
\end{equation}
We now exploit the Seiberg-Witten Map \cite{sw,jur} to revert back
to the ordinary  spacetime degrees of freedom. The explicit
identifications between NC and ordinary spacetime counterparts of
the fields, to the lowest non-trivial order in $\theta$ are,
$$
\hat A_{\mu}=A_{\mu}+\theta^{\sigma\rho}A_{\rho}(\partial_{\sigma}
A_{\mu}-\frac{1}{2} \partial_{\mu} A_{\sigma})
$$
\begin{equation}
\hat \phi =\phi -\frac{1}{2}\theta^{\rho\sigma} A_{\rho}\partial_{\sigma}\phi ~;~~
\hat \lambda = \lambda -\frac{1}{2}\theta^{\rho\sigma} A_{\rho}\partial_{\sigma} \lambda .
\label{swm}
\end{equation}
As stated before, the "`hatted"' variables on the left are NC
degrees of  freedom and gauge transformation parameter. The higher
order terms in $\theta$ are kept out of contention as there are
certain non-uniqueness involved in the $O(\theta^{2})$ mapping.
The significance of the Seiberg-Witten map is that under an NC or
$*$-gauge transformation of $\hat A_{\mu}$ by,
$$\hat\delta \hat A_{\mu}=\partial_{\mu}\hat \lambda +i[\hat\lambda ,\hat A_{\mu}]_{*},$$
$A_{\mu}$ will undergo the transformation $$\delta A_\mu
=\partial_{\mu}\lambda . $$ Subsequently, under this mapping, a
gauge invariant object in conventional spacetime will be mapped to
its NC counterpart, which will be $*$-gauge invariant. This is
crucial as it ensures that the ordinary spacetime action that  we
recover from the NC action (\ref{ncac}) by applying the
Seiberg-Witen Map will be gauge invariant. Thus the NC action
(\ref{ncac}) in ordinary spacetime variables reads,
\begin{equation}
\hat S =\int d^{3}x [( D^{\mu}\phi )^{*} D_{\mu}\phi +
\frac{1}{2}\theta^{\alpha\beta} \{F_{\alpha\mu}((D_{\beta}\phi
)^{*} D^{\mu}\phi + (D^{\mu}\phi )^{*} D_{\beta}\phi
)-\frac{1}{2}F_{\alpha\beta}(D^{\mu}\phi )^{*} D_{\mu}\phi \}]
\label{ncac1}
\end{equation}
The above action is manifestly gauge invariant.
Remember that so far we have not introduced the $CP^{1}$ target space constraint
in the NC spacetime setup. {\it{Let us assume the constraint to be identical to the
ordinary spacetime one }} \cite{sg}, i.e.,
\begin{equation}
\phi ^{*}\phi =1.
\label{cp}
\end{equation}
The reasoning is as follows \cite{sg}. Basically, after utilizing the
Seiberg-Witten Map,  we have returned to the ordinary spacetime
and its associated dynamical variables and the effects of
noncommutativity appears only as additional interaction terms in
the action. Hence it is natural to keep the $CP^{1}$ constraint
unchanged. (For more details, see \cite{sg}.)

This allows us to write,
\begin{equation}
A_{\mu}=-i\phi^{*}\partial_{\mu}\phi +a_{\mu}(\theta)
\label{nca}
\end{equation}
with $a_{\mu}$ denoting the $O(\theta )$ correction,  obtained
from (\ref{ncac1},\ref{cp}). For $\theta =0$. $A_{\mu}$ reduces to its original form.
Note that $a_{\mu}$ is {\it {gauge invariant}}.
Thus the $U(1)$ gauge transformation of  $A_{\mu}$  remains
intact, at least to  $O(\theta)$.
 Keeping in mind the constraint $\phi^{*}\phi =1$,
 let us now substitute (\ref{nca}) in the NC action (\ref{ncac1}).
 Since we are concerned only with the $O(\theta)$ correction, in
 the $\theta$-term of the action, we can use $A_{\mu}=-i\phi^{*}\partial_{\mu}\phi$.
 However, in the first term in the action,    we must incorporate the full expression
 for $A_{\mu}$ given in (\ref{nca}). Remarkably, the constraint condition conspire to
 cancel the effect of the $O(\theta)$ correction term $a_{\mu}$.  Finally it boils down
 to the following: the action for the  NC $CP^{1}$ model to $O(\theta )$ is given by
 (\ref{ncac1}) with the {\it{identifications}} $A_{\mu}=-i\phi^{*}\partial_{\mu}\phi ,~F_{\mu\nu}=\partial_{\mu}A_\nu -\partial_\nu A_\mu$ and the {\it{constraint}}
 $\phi^{*}\phi =1$ \cite{sg}.

\vskip .5cm
\begin{center}
{\bf{Section III - Energy-momentum tensor for the NC $CP(1)$ model}}
\end{center}
As is well-known, in general, it is not possible to obtain a
symmetric, gauge invariant and conserved EM tensor in an NC field
theory, with noncommutativity of the form of (\ref{nc}). There are
two forms of EM tensor in vogue \cite{grim}: a manifestly
symmetric form, obtained from the variation of the action with
respect to the metric, and the canonical form, following the
Noether prescription. The former is covariantly conserved whereas
the latter is conserved. Since we are interested in the space-time
invariance properties and Poincare generators, we will concentrate
on the canonical (Noether) form, which is given by,

\begin{equation}
T_{\mu\nu}=\frac{\delta \hat L}{\delta (\partial^{\mu}\phi ^{*})}\partial_{\nu}\phi^{*}
+\frac{\delta \hat L}{\delta (\partial^{\mu}\phi )}\partial_{\nu}\phi  -g_{\mu\nu}\hat L .
\label{0}
\end{equation}
In the present case, for the action (\ref{ncac1}),
$$
\hat S =\int d^{3}x [( D^{\mu}\phi )^{*} D_{\mu}\phi +
\frac{1}{2}\theta^{\alpha\beta} \{F_{\alpha\mu}((D_{\beta}\phi
)^{*} D^{\mu}\phi + (D^{\mu}\phi )^{*} D_{\beta}\phi
)-\frac{1}{2}F_{\alpha\beta}(D^{\mu}\phi )^{*} D_{\mu}\phi \}
$$
\begin{equation}
+\Lambda (\phi^{*}\phi -1)]
\label{ncac2}
\end{equation}
the EM tensor $T_{\mu\nu}$ is,
$$
T_{\mu\nu}=[D_{\mu}\phi^{*}D_{\nu}\phi +(\mu\Leftrightarrow \nu )](1-\frac{1}{4}\theta^{\alpha\beta} F_{\alpha\beta})+\frac{1}{2}\theta ^{\alpha\beta}[F_{\alpha\nu}(D_{\beta}\phi^{*}D_{\mu}\phi +D_{\mu}\phi^{*}D_{\beta}\phi )+(\mu\Leftrightarrow \nu )]$$
\begin{equation}
-\frac{1}{2}\theta_{\alpha\mu}F^{\alpha}_{~\nu}\mid D\phi\mid ^2+\frac{1}{2}\theta_{\mu\beta}[F_{\nu\alpha}(D^{\beta}\phi^{*}D^{\alpha}\phi + D^{\alpha}\phi^{*}D^{\beta}\phi )+F^{\alpha\beta}(D_{\alpha}\phi^{*}D_{\nu}\phi + D_{\nu}\phi^{*}D_{\alpha}\phi )].
\label{2}
\end{equation}
 In order to obtain a symmetric $T_{\mu\nu}$, at least for $\theta =0$, we have
 rewritten the covariant derivatives in the following equivalent form,
$$
(D^{\mu}\phi )_{a}=\partial^{\mu}\phi_{a}-(\phi^{*}\partial^{\mu}\phi )\phi_{a}
=\partial^{\mu}\phi_{a}-\frac{1}{2}(\phi^{*}\partial^{\mu}\phi -\partial^{\mu}\phi^{*}\phi ) \phi_{a},$$
\begin{equation}
(D^{\mu}\phi )^{*}_{a}=\partial^{\mu}\phi^{*}_{a}+(\phi^{*}\partial^{\mu}\phi )\phi_{a}
=\partial^{\mu}\phi^{*}_{a}+\frac{1}{2}(\phi^{*}\partial^{\mu}\phi -\partial^{\mu}\phi^{*}\phi ) \phi^{*}_{a}.
\label{1}
\end{equation}
Note that $\mu\Leftrightarrow \nu$  symmetry of $T_{\mu\nu}$ is
destroyed by some of the $\theta$-terms.  However, $T_{\mu\nu}$ is
 manifestly gauge
invariant and conserved (as it is derived  from the canonical definition (\ref{0})).
\vskip .5cm
\begin{center}
{\bf{Section IV - Hamiltonian formulation and constraint analysis}}
\end{center}
 \vskip .5cm
 Let us now perform a Hamiltonian constraint analysis, in the Dirac \cite{di} scheme,
 which entails in obtaining the full set of constraints in a given theory. Furthermore,
 the constraints are classified in to the largest set of commuting constraints (the First
 Class Constraints (FCC)) and the remaining non-commuting constraints (Second Class
 Constraints (SCC)). The presence of FCCs indicate local gauge invariances.
 For a consistent quantization programme, the SCCs are taken into account by
 replacing the Poisson brackets by Dirac brackets \cite{di}, that is defined
 below for two generic variables $A$ and $B$,
\begin{equation}
\{A,B\}_{DB}=\{A,B\}-\{A,\chi _i\}\{\chi _i,\chi _j\}^{-1}\{\chi _j,B\},
\label{db}
\end{equation}
In (\ref{db}), Poisson brackets are used in the right hand side and $\{\chi _i,\chi _j\}^{-1}$
denotes the inverse of the constraint Poisson bracket matrix $\{\chi _i,\chi _j\}$, the latter
being invertible for SCC $\chi_i$. Notice that using Dirac brackets allows us to put the SCCs
strongly equal to zero since they commute with everything in the Dirac bracket sense,
$$\{A,\chi_{i}\}_{DB}=\{\chi_{j},B\}_{DB}=0.$$

  We closely follow the earlier work \cite{vij} on $CP(1)$ model in
  ordinary spacetime. {\footnote{In an alternative extended space quantization scheme \cite{bt},
   the $CP(1)$ model has been discussed in \cite{cha}.}} Only spatial noncommutativity is being
   considered here, that is $\theta^{01}=\theta^{02}=0,~ \theta^{12}=\theta\epsilon^{12}$.
  The canonically conjugate momenta, as obtained from the action (\ref{ncac2}), are
$$
\pi_{a}=(1+C)D^0\phi^{*}_{a}-i\theta\epsilon^{ij}(D^0\phi^{*}D^j\phi)D^i\phi^{*}_{a},$$
\begin{equation}
\pi^{*}_{a}=(1+C)D^0\phi_{a}+i\theta\epsilon^{ij}(D^0\phi D^j\phi^{*})D^i\phi_{a},
\label{3}
\end{equation}
where $C\equiv -\frac{1}{4}\theta\epsilon^{ij}F^{ij}$. We
immediately find the following two primary constraints, $\psi_{2},
~\psi_{3}$, along with the $CP(1)$ primary constraint $\psi_{1}$,
\begin{equation}
\psi_{1}\equiv \phi^{*}\phi -1\approx 0~;~\psi_{2}\equiv \phi\pi \approx 0~;~\psi_{3}\equiv \phi^{*}\pi^{*} \approx 0.
\label{9}
\end{equation}
Using the basic canonical Poisson brackets,
\begin{equation}
\{\phi_{a}(x),\pi_{b}(y)\}=\delta_{ab}\delta (x-y)~;~\{\phi^{*}_{a}(x),\pi^{*}_{b}(y)\}=\delta_{ab}\delta (x-y)
\label{10}
\end{equation}
it is revealed that the linear combination
\begin{equation}
\xi \equiv \phi\pi -\phi^{*}\pi^{*}\approx 0
\label{11}
\end{equation}
commutes with the other two constraints,
\begin{equation}
\chi_{1}\equiv \phi^{*}\phi -1=0~;~\chi_{2}\equiv \phi\pi +\phi^{*}\pi^{*}=0.
\label{12}
\end{equation}
However, the non-vanishing bracket,
\begin{equation}
\{\chi_{1}(x),\chi_{2}(y)\}=2(\phi^{*}(x)\phi (x))\delta (x-y)\approx 2\delta (x-y),
\label{scc}
\end{equation}
indicates that the above pair of constraints are Second Class
Constraints (SCC) \cite{di}.  In the present case the basic Dirac
brackets are computed below,
$$\{\phi_{a}(x),\phi_{b}(y)\}=\{\phi_{a}(x),\phi^{*}_{b}(y)\}=0,$$
$$
\{\phi_{a}(x),\pi_{b}(y)\}=(\delta_{ab}-\frac{1}{2}\phi_{a}\phi^{*}_{b})\delta (x-y)~;~
\{\phi_{a}(x),\pi^{*}_{b}(y)\}=-\frac{1}{2}\phi_{a}\phi_{b}\delta (x-y),$$
\begin{equation}
\{\pi_{a}(x),\pi_{b}(y)\}=\frac{1}{2}(\pi_{a}\phi^{*}_{b}-\pi_{b}\phi^{*}_{a})\delta (x-y)~;~
\{\pi_{a}(x),\pi^{*}_{b}(y)\}=\frac{1}{2}(\pi_{a}\phi_{b}-\pi^{*}_{b}\phi^{*}_{a})\delta (x-y).
\label{8}
\end{equation}
Since in the subsequent analysis only Dirac brackets are used, we have avoided the notation $\{,\}_{DB}$. Complex conjugation reproduces rest of the Dirac brackets. From now on we will be utilising Dirac brackets and exploit the SCC relations strongly, that is $\chi_{1}=\chi_{2}=0$.

In order to find the full set of constraints, we now compute the
canonical Hamiltonian. This means that the time derivatives are to
be expressed  in terms of the phase space variables. This is
carried out to $O(\theta )$. To that end, we first note from
(\ref{3}) that
\begin{equation}
D^0\phi^{*}_{a}=\pi_{a}+O(\theta)~,~~D^0\phi_{a}=\pi^{*}_{a}+O(\theta).
\label{5}
\end{equation}
This yields the following relations,
$$
\pi_{a}=(1+C)D^0\phi^{*}_{a}-i\theta\epsilon^{ij}(\pi D^j\phi )D^i\phi^{*}_{a}+O(\theta^{2}),$$
\begin{equation}
\pi^{*}_{a}=(1+C)D^0\phi_{a}+i\theta\epsilon^{ij}(\pi^{*} D^j\phi ^{*} )D^i\phi_{a}+O(\theta^{2}).
\label{6}
\end{equation}
The above equations allow us to rewrite the time derivatives as,
$$D^0\phi_{a}
=(1-C)\pi^{*}_{a}-i\theta\epsilon^{ij}(\pi^{*} D^j\phi^{*} )D^i\phi_{a}+O(\theta^{2}),$$
\begin{equation}
D^0\phi^{*}_{a}
=(1-C)\pi_{a}+i\theta\epsilon^{ij}(\pi D^j\phi )D^i\phi^{*}_{a}+O(\theta^{2}).
\label{7}
\end{equation}
This leads us to the canonical Hamiltonian, (with noncommutative effects up to $O(\theta )$),
$$T_{00}=2(1+C)D_0\phi^{*}D_0\phi +\theta\epsilon_{ij}F_{i0}(D_0\phi^{*}D_j\phi
 +D_j\phi^{*}D_0\phi)-{\cal{L}}$$
\begin{equation}
=(\pi^{*}\pi +D^{k}\phi^{*}D^{k}\phi )(1-C)+i\theta\epsilon^{ij}(\pi^{*}D^{i}\phi^{*})(\pi D^{j}\phi ).
\label{14}
\end{equation}
The total Hamiltonian, in the terminology of Dirac \cite{di} is
\begin{equation}
{\cal{H}}=T_{00}+\Lambda (x)\xi (x).
\label{toth}
\end{equation}
It is now straightforward (though tedious) to check explicitly that
\begin{equation}
\{\xi (x),H\}=\{\xi (x),\int d^2y~ T_{00}(y)\}=0.
\label{15}
\end{equation}
This demonstration is very significant as it shows that there are
no further constraints and that $\xi \equiv \phi\pi
-\phi^{*}\pi^{*}\approx 0$ is the only First Class Constraint
(FCC) \cite{di}. The presence of the FCC signals a gauge
invariance - $U(1)$ in the present case. This fact was apparent in
the explicit form of the action (\ref{ncac2}) as well. It is
trivial to ensure that the FCC $\xi (x) $ functions properly as
the generator $Q$ of $U(1)$ gauge transformation $(Q\equiv \int
d^2x~\alpha (x)\xi (x))$ by evaluating the brackets,
$$
\delta_{\alpha}\phi_{a}(x)\equiv \{\phi_{a}(x),Q\}=\alpha (x)\phi_{a}(x)~;~
\delta_{\alpha}\phi^{*}_{a}(x)\equiv \{\phi^{*}_{a}(x),Q\}=-\alpha (x)\phi^{*}_{a}(x),$$
\begin{equation}
\delta_{\alpha}\pi_{a}(x)\equiv \{\pi_{a}(x),Q\}=-\alpha (x)\pi_{a}(x)~;~
\delta_{\alpha}\pi^{*}_{a}(x)\equiv \{\pi^{*}_{a}(x),Q\}=\alpha (x)\pi^{*}_{a}(x),
\label{gauge}
\end{equation}
where $\alpha (x)$ denotes the infinitesimal gauge transformation parameter.
\vskip .5cm
\begin{center}
{\bf{Section V: Schwinger condition, Poincare algebra and the equation of motion }}
\end{center}
 \vskip .5cm
 Finally we are ready to tackle the question of the exact nature of Lorentz symmetry
 violation induced by noncommutativity. We will
closely follow  the conventional field theoretic approach in
ordinary spacetime \cite{vij}.

 First of all, let us now compute the spatial momenta
 \begin{equation}
 P_i=\int d^2x~ T_{0i}(x),
 \label{mom}
 \end{equation}
where $T_{\mu\nu}$ in (\ref{2}) reproduces
$$T_{0i}=(D_0\phi^{*}D_i\phi +D_i\phi^{*}D_0\phi )(1+C)$$
$$+\frac{1}{2}\theta\epsilon_{jk}[F_{ji}(D_0\phi^{*}D_k\phi
 +D_k\phi^{*}D_0\phi)+F_{j0}(D_k\phi^{*}D_i\phi +D_i\phi^{*}D_k\phi)]$$
\begin{equation}
=\pi D_i\phi +\pi^{*} D_i\phi^{*}=\pi \partial_i\phi +\pi^{*} \partial_i\phi^{*}+\xi(x) \phi^{*}\partial_{i}\phi \approx \pi \partial_i\phi +\pi^{*} \partial_i\phi^{*}.
\label{13}
\end{equation}
We immediately find that $P_i$ generates correct transformations
among the degrees of freedom:
\begin{equation}
\{\phi (x),P_{i}\}=\{\phi (x),\int d^2y ~T_{0i}(y)\}=\partial_{i}\phi (x)~;~\{\pi (x) (x),P_{i}\}=\{\pi (x),\int d^2y ~T_{0i}(y)\}=\partial_{i}\pi (x),
\label{16}
\end{equation}
where the Dirac brackets in (\ref{8}) are used. Next we show that
the Hamiltonian (\ref{14}) represents the generator of time
translation,
$$\{\phi_{a}(x),H\}\equiv \{\phi_{a}(x),\int d^2x~T_{00}(x)\}$$
\begin{equation}
=\pi^{*}_{a}+i\theta\epsilon^{ij}[(\pi^{*}D^{i}\phi^{*})D^{j}\phi_{a}-\frac{1}{2}(D^{i}\phi^{*}D^{j}\phi )\pi^{*}_{a}]=D^0\phi_{a},
\label{time}
\end{equation}
where we have used (\ref{6}) to replace the momenta. Now notice an interesting fact that
$$\phi^{*}\partial_{0}\phi =\frac{1}{2}(\phi^{*}\partial_{0}\phi -\phi\partial_{0}\phi^{*})$$
\begin{equation}
=-\frac{1}{2}\xi (1-i\theta\epsilon^{ij}D^i\phi^*D^j\phi ).
\label{ti}
\end{equation}
Putting (\ref{ti}) back in (\ref{time}) this means that modulo the FCC $\xi (x)$,
\begin{equation}
\{\phi_{a}(x),H\}=\partial_{0}\phi_{a},
\label{tt}
\end{equation}
so that the correct Hamiltonian equation of motion is reproduced.
The above results confirm the conservation of the energy and
momenta, thereby ensuring Observer Lorentz invariance \cite{har}.
Obviously this is expected since the action is manifestly
translation invariant, but the cancellation of the $\theta$-term
is indeed non-trivial. However, we emphasize that this is the
first explicit demonstration of the conservation principle in a
particular noncommutative field theory model, in the Hamiltonian
framework. This one of our main results.

We now consider how far it is possible to construct a Lorentz
covariant equation of motion for $\phi_{a}$. Let us start with the
ordinary spacetime $CP(1)$ model ($\theta =0$). The Hamiltonian
equations of motion yields,
\begin{equation}
\partial_{0}\phi_{a}=\pi_{a}^{*}~;~\partial_{0}\pi_{a}^{*}=-(\pi^{*}\pi )\phi_{a} +D^k(D^k\phi_{a}).
\label{eqm}
\end{equation}
Once again exploiting the previous argument leading to (\ref{ti}),
we can express (\ref{eqm}) in a manifestly covariant form, (modulo
the FCC $\xi$),
\begin{equation}
D^\mu (D_\mu\phi_{a})=-(\pi^{*}\pi )\phi_{a}.
\label{eqm1}
\end{equation}
The $O(\theta)$ correction to the above equation is
straightforward to compute but the the explicit form is quite
involved and not particularly illuminating. Writing the
$O(\theta)$ equation of motion schematically,
\begin{equation}
D^\mu (D_\mu\phi_{a})=-(\pi^{*}\pi )\phi_{a} +\epsilon^{\mu\nu\lambda}\theta_{\mu}A_{\nu\lambda},
\label{eqm2}
\end{equation}
we note that $A_{\nu\lambda}$ contains manifestly Lorentz
covariant and non-covariant terms, comprising of field variables.
Hence one is not able to recover a fully Lorentz covariant
dynamical equation of motion in NC field theory.

Our next objective is the study of the Schwinger condition \cite{schw}, the
validity of which is necessary and sufficient to guarantee
Poincare invariance. This requires verification of the all
important bracket \cite{schw},
\begin{equation}
\{T_{00}(x),T_{0i}(y)\}=(T^{0i}(x)+T^{0i}(y))\partial^{i}_{(x)}\delta (x-y),
\label{17a}
\end{equation}
known as the Schwinger condition. In the present case, after a long calculation, we find,
\begin{equation}
\{T_{00}(x),T_{0i}(y)\}=(T^{0i}(x)+T^{0i}(y))\partial^{i}_{(x)}\delta (x-y)+(\tau^{i}(x)+\tau^{i}(y))\partial^{i}_{(x)}\delta (x-y),
\label{17}
\end{equation}
where
$$
\tau^{i}=i\theta \epsilon^{ij}[\frac{1}{2}(\pi^{*}\pi +D^k\phi^{*}D^k\phi)(\pi D^j\phi -\pi ^*D^j\phi ^*)$$
\begin{equation}
+(\pi^{*}D^k\phi^{*})(D^j\phi^{*}D^k\phi)-(\pi D^k\phi )(D^k\phi^{*}D^j\phi) ].
\label{18}
\end{equation}
 Clearly in the present instance, the Schwinger condition is not satisfied
 because  $\tau_i\neq 0$ in general, indicating a violation of Particle Lorentz
 symmetry \cite{har}. The reason is clearly the introduction of  a constant tensor,
 in the form of the noncommutativity parameter $\theta_{\mu\nu}$, which singles out a
 fixed direction in spacetime. This is the other major result that we had advertised.

It is interesting to note that, in 2+1-dimensions, even for
$\theta\neq 0$, Schwinger pcondition and subsequently Poincare
invariance can be maintained in this model provided the fields are
such that $\tau_i$ vanishes identically. This type of scenario has
not been reported before.

If we define the Poincare generators in the conventional way,
\begin{equation}
P_{\mu}\equiv \int d^2x ~T_{0\mu}~;~J_{\mu\nu}\equiv \int d^2x~ (x_{\mu}T_{0\nu}-x_{\nu}T_{0\mu}),
\label{poin}
\end{equation}
the sector of the Poincare algebra involving the Lorentz boost
generators $J_{0i}$ will be violated. It is not surprising that,
although we have introduced spatial noncommutativity, still the
angular momentum algebra remains intact, indicating rotational
invariance. This is because the asymmetry (via noncommutativity)
is actually introduced in the time direction, which is easily seen
if we look at  $\theta^{\mu}\equiv
\frac{1}{2}\epsilon^{\mu\nu\lambda}\theta_{\nu\lambda}$, the dual
of $\theta_{\mu\nu}$. In the present example, a non-zero
$\theta_{12}$ yield a non-zero $\theta^{0}$.

\vskip .5cm
\begin{center}
{\bf{Section VI: Conclusions and Future Prospects}}
\end{center}
 \vskip .5cm
Let us summarize our work. In \cite{sg} we had provided an
alternative formulation of the noncommutative extension of the
$CP(1)$ model, that was distinct from the existing ones
\cite{nccp}. The present work deals with the Hamiltonian
formulation of the model of \cite{sg}. We emphasize that probably
this is the first instance where a noncommutative field theory has
been studied in the Hamiltonian framework and most of the basic
observations will be relevant for the study of a generic
noncommutative field theory in Hamiltonian framework.

The aim is to study in detail the characteristic features of the
violation of Lorentz invariance in a noncommutative field theory.
This subtle issue first appeared in \cite{har}, where it was
pointed out that two distinct types of Lorentz transformations are
to considered: the Observer and the Particle Lorentz
transformations. Symmetry under the former is maintained (in
noncommutative theories), due to the translation invariance of the
theory, thereby leading to the conservation of energy and
momentum. The latter is destroyed owing to the presence of the
constant tensor $\theta_{\mu\nu}$.

We have successfully addressed the above issues in the present
work. We have constructed the Hamiltonian and momentum operators
and have shown that they act properly as the  generators  time and
space translations. This is related to the Observer Lorentz
invariance. On the other hand, we have shown that Schwinger
condition and subsequently the Poincare algebra is not respected
and that one can not derive a Lorentz covariant dynamical field
equation. These features signal a loss of the Particle Lorentz
symmetry.

The age old Hamiltonian constraint analysis, as formulated by
Dirac \cite{di}, has been instrumental in our analysis. The model
contains both First Class and Second Class constraints, conforming
to the classification scheme of Dirac \cite{di}. The Dirac
brackets \cite{di} have been computed and they are exploited
throughout in arriving at the above mentioned results.

An intriguing open problem is the introduction of the Hopf term in
the $CP(1)$ model and its subsequent noncommutative extension. The
classic work of \cite{wz} was the demonstration that the Hopf term
is able to impart fractional spin and statistics to the solitons
of the Non-linear $\sigma $-model. This was further corroborated
in \cite{vij} in a Hamiltonian formulation. The advantage in an
alternative $CP(1)$ representation of the Non-linear $\sigma
$-model is that the Hopf term is reduced to a local expression
\cite{wuz} in terms of $CP(1)$ degrees of freedom. One of our
future projects is to study the effects of noncommutativity on the
anyons induced by the Hopf term.

\end{document}